\documentclass[APS, preprint,superscriptaddress]{revtex4-2}
\usepackage{amsmath}
\usepackage{bm}
\usepackage{physics}

\usepackage{array}
\usepackage{dcolumn}
\usepackage{adjustbox}
\newcolumntype{C}[1]{>{\centering\arraybackslash}m{#1}}

\usepackage{graphicx}
\usepackage{rotating}

\usepackage[caption=false]{subfig}

\usepackage{setspace}

\usepackage{placeins}

\begin{document}

\preprint{APS/123-QED}

\title{Stochastic GW with the Orthogonalized Projector Augmented Wave Method}


\author{Dimitri Bazile}
\affiliation{Department of Chemistry and Biochemistry, University of California, Los Angeles, Los Angeles, CA 90095, USA}

\author{Minh Nguyen}
\affiliation{Center for Computing Research, Sandia National Laboratories, Albuquerque, NM 87123, USA}

\author{Yuji Kon}
\affiliation{Department of Chemistry and Biochemistry, University of California, Los Angeles, Los Angeles, CA 90095, USA}

\author{Tucker Allen}
\affiliation{Department of Chemistry and Biochemistry, University of California, Los Angeles, Los Angeles, CA 90095, USA}

\author{Daniel Neuhauser}
\email{dxn@ucla.edu}
\affiliation{Department of Chemistry and Biochemistry, University of California, Los Angeles, Los Angeles, CA 90095, USA}

\date{\today}

\begin{abstract}
We introduce stochastic GW with the orthogonalized projector augmented-wave method (OPAW-sGW).  This implementation enables accurate quasiparticle band gaps on significantly coarser real-space grids than norm-conserving pseudopotential sGW (NCPP-sGW). The orthogonalized PAW representation preserves the formal all-electron character and enables stochastic sampling of the Green’s function and screened Coulomb interaction.
\end{abstract}
\maketitle

\section{Introduction}
Density functional theory (DFT) successfully predicts many ground-state properties of molecular and solid-state systems; however, it often provides an inaccurate description of excited-state phenomena. In particular,  DFT  often underestimates band gaps and frequently fails to reproduce optical spectra \cite{Perdew2009}.   The GW approximation within the framework of many-body perturbation theory (MBPT) provides a systematic correction to these deficiencies and has become a reliable method for computing quasiparticle (QP) energies \cite{Hedin1965,Aryasetiawan1998}. GW also forms the basis for modeling optical excitations through the Bethe–Salpeter equation (BSE), which explicitly accounts for electron–hole interactions \cite{Blase2020}.

In GW, exchange and correlation effects beyond mean-field theory are described by the single-particle self-energy,
$\Sigma(t) = iG(t)W(t)$, where $G$ is the single-particle Green's function and $W$ is the screened Coulomb interaction \cite{Golze2019}.  Several approximations exist for evaluating the self-energy. In the one-shot $G_0W_0$ approach, $G_0$ and $W_0$ are constructed from mean-field orbitals and energies obtained from either DFT or Hartree--Fock (HF), and neither quantity is updated self-consistently \cite{Stan2009,Vlek2018_2}. In partially eigenvalue self-consistent GW, often denoted $GW_0$ or ev-$GW$, the eigenvalues entering $G$ are updated self-consistently while the screened interaction $W$ remains fixed \cite{Stan2009}. In fully self-consistent GW (scGW), both $G$ and $W$ are iteratively updated until convergence \cite{Holm1998,PhysRevB.106.235104,Wen2024}. Several electronic-structure packages that implement these GW methods include ABINIT, MOLGW, BerkeleyGW, VASP, Exciting, and Yambo \cite{Romero2020,Bruneval2016,Deslippe2012,Shishkin2006,Gulans2014,Marini2009}. These codes compute quasiparticle spectra, band gaps, and, depending on the target system, electronic band structures.

The GW methods mentioned above typically scale quartically \(\mathcal{O}(N^4)\)with the size of the system $N$, making them challenging for large molecules or periodic solids \cite{Bintrim2021, Wilhelm2021}. This steep scaling arises from the need to compute a non-Hermitian, spatially nonlocal self-energy, $\Sigma(\omega = \epsilon_n^{\mathrm{QP}})$. Modern codes reduce this scaling by compressing band summations within plane-wave frameworks using pseudobands \cite{PhysRevLett.132.086401,Zhang2025}, by employing Gaussian basis sets and/or the resolution-of-the-identity technique, as implemented, for example, in FIESTA \cite{Frster2023}, or by using atomic density fitting \cite{Frster2020}. These approaches reduce the scaling from quartic \(\mathcal{O}(N^4)\) to cubic \(\mathcal{O}(N^3)\).

An alternative approach is stochastic GW (sGW), in which the electronic self-energy is evaluated by stochastic sampling of the Green’s function, \(G\), and the screened Coulomb interaction, \(W\)  \cite{Neuhauser2014, Vlek2017}.  In this approach, explicit summations over large numbers of occupied and unoccupied states are replaced by a much smaller set of stochastic orbitals, which are random linear combinations of occupied and unoccupied states. The stochastic techniques used in this approach dramatically reduce the computational cost from high-order polynomial scaling to linear \(\mathcal{O}(N)\) or sub-quadratic \(\mathcal{O}(N^2)\) \cite{Vlek2018, Allen2024, Weng2021}, with the potential for further time reduction through massively parallel GPU acceleration~\cite{PHIL2026}. As a result, sGW enables accurate QP energy calculations for systems far larger than those accessible with deterministic GW methods \cite{Allen2024}. 

Previous implementations of sGW used norm-conserving pseudopotentials (NCPPs), which are smooth effective potentials that act on valence electrons while mimicking the strong Coulomb potential of the nucleus and core electrons. NCPPs reduce computational cost and complexity by eliminating the need to explicitly treat core-electron interactions or core states \cite{Hamann1979,Troullier1991}. However, they typically require fine real-space grids, or equivalently, large plane-wave kinetic-energy cutoffs, to accurately represent the rapid oscillations near the nucleus.

The projector augmented-wave (PAW) method, developed by Bl\"ochl, Kresse, and Joubert \cite{Blchl1994,Kresse1999}, maps all-electron wavefunctions onto smooth pseudo-wavefunctions through a linear transformation, enabling efficient plane-wave representations.  In this formalism, pseudo-wavefunctions are expanded in plane waves across the simulation cell, while atom-centered augmentation functions near each nucleus reconstruct the all-electron wavefunctions and operator matrix elements. This reconstruction restores the all-electron character near the nucleus, allowing PAW to accurately model core-level spectra and NMR chemical shifts in molecules \cite{Charpentier2011,Ljungberg2011}.  Other all-electron approaches suitable for transition-metal complexes and correlated materials include the augmented plane-wave (APW) method and its variants, such as linearized APW (LAPW) and fully linearized APW \cite{Dupuy}. 

PAW has also been combined with several post-DFT methods, including the Bethe–Salpeter equation (BSE) \cite{Yan2012}, time-dependent density functional theory (TDDFT) \cite{Walter2008}, and GW approaches \cite{Shishkin2006,Hser2013}.  Combining PAW with sGW, however, presents a major challenge: the stochastic techniques used in sGW require an orthogonal basis, whereas the PAW basis is nonorthogonal. To address this issue, we recently developed the orthogonalized projector augmented-wave (OPAW) method, which orthogonalizes PAW wavefunctions while preserving the essential features of the original PAW representation. OPAW has so far been applied to static DFT and real-time TDDFT \cite{Li2020,Nguyen2024}. Extending OPAW to sGW therefore represents a natural next step toward improving the accuracy of charged excitations at reduced computational cost.

In this article, we benchmark OPAW-sGW against NCPP-sGW for a diverse set of molecular systems, including planar and curved conjugated hydrocarbons, a donor–acceptor complex, and photosynthetic chromophores. The results show that OPAW-sGW achieves the same accuracy as NCPP-sGW while using significantly coarser real-space grids, which substantially reduces memory requirements. The theory section presents the sGW formalism within the OPAW representation. In the results section, we compare OPAW-sGW and NCPP-sGW through grid-spacing convergence tests for naphthalene and through comparisons of QP band gaps and self-energy contributions for larger systems. All reported sGW band gaps include the simplified eigenvalue self-consistent correction, $\bar{\Delta}GW_0$, applied to the s$G_0W_0$ results following Ref. \cite{Vlek2018_2}. This rigid, scissors-like eigenvalue self-consistency improves sGW band gaps at essentially no additional computational cost. We also discuss the computational speedups obtained with OPAW-sGW. Finally, we summarize the main findings and outline future directions.

\section{Theory}
\subsection{PAW and OPAW Formalism}
Within the PAW formalism, the all-electron (AE) wavefunctions, $\psi_n$, are constructed from a smooth auxiliary (pseudo) wavefunctions $\tilde{\psi}_n$ through the linear transformation operator, $\hat{T}$:
\begin{equation}
|\psi_n \rangle = \hat{T} |\tilde{\psi}_n \rangle
\equiv
|\tilde{\psi}_n \rangle
+ \sum_{a,i} \left( |\phi_i^{(a)} \rangle - |\tilde{\phi}_i^{(a)} \rangle \right)
\langle p_i^{(a)} | \tilde{\psi}_n \rangle .
\label{eq:paw}
\end{equation}
Here, $a$ denotes the atomic index and $i$ labels the partial-wave channels associated with each atom. The functions $\phi_i^{(a)}$ and $\tilde{\phi}_i^{(a)}$ represent the true atomic partial waves and their smooth counterparts, respectively. The projector functions $p_i^{(a)}$ span the Hilbert space within the augmentation spheres.

Using the auxiliary wavefunctions, the Kohn–Sham (KS) eigenvalue problem becomes a generalized eigenvalue equation:
\begin{equation}
\tilde{H} |\tilde{\psi}_n \rangle = \epsilon^{KS}_n \hat{S} |\tilde{\psi}_n \rangle ,
\label{eq:paw_eq}
\end{equation}
where $\tilde{H}$ is the PAW Hamiltonian and $\hat{S}$ is the overlap operator,
\begin{equation}
\hat{S} = \hat{T}^\dagger \hat{T}.
\label{eq:overlap}
\end{equation}
The overlap operator ensures orthonormality of the all-electron wavefunctions,
\begin{equation}
\langle \psi_n | \psi_m \rangle
= \langle \tilde{\psi}_n | \hat{S} | \tilde{\psi}_m \rangle
= \delta_{nm}.
\label{eq:S_op}
\end{equation}

In the Kresse–Joubert formulation, the PAW Hamiltonian is
\begin{equation}
\tilde{H} 
= \hat{T}_{KE} + \tilde{v}_{eff} + \hat{D},
\label{eq:paw_ham}
\end{equation}
where $\hat{T}_{KE}$ is the kinetic energy operator, $\tilde{v}_{eff}$ is an effective one-electron potential, and $\hat{D}$ is a nonlocal operator arising from the transformation operator $\hat{T}$. Further details on these terms can be found in Refs.~\cite{Torrent2008,Torrent2024PAW}.

To obtain an orthonormal representation of the PAW Hamiltonian and eigenstates, we apply the transformations
\begin{equation}
\bar{H} = \hat{S}^{-1/2} \tilde{H} \hat{S}^{-1/2},
\label{eq:opaw_ham}
\end{equation}

\begin{equation}
|\bar{\psi}_n \rangle = \hat{S}^{1/2} |\tilde{\psi}_n \rangle .
\label{opaw_state}
\end{equation}
Here, $\bar{H}$ and $|\bar{\psi}_n\rangle$ denote the OPAW Hamiltonian and wavefunctions, respectively. This transformation yields the standard Hermitian eigenvalue problem
\begin{equation}
\bar{H} |\bar{\psi}_n \rangle = \epsilon_n^{\mathrm{KS}} |\bar{\psi}_n \rangle .
\label{eq:opaw_eq}
\end{equation}
The procedure for computing arbitrary powers of $\hat{S}$ is straightforward and has been described in detail in our previous work \cite{Li2020}.

\subsection{OPAW with Stochastic GW}
Single-shot GW corrects KS energies by replacing the exchange–correlation potential $v_{xc}$ with a nonlocal, frequency-dependent self-energy operator $\Sigma$. Within the diagonal approximation, quasiparticle (QP) energies are obtained from the approximate Dyson equation
\begin{equation}
\epsilon_n^{QP}
=
\epsilon_n^{KS}
+
\left\langle \bar{\psi}_n \left|
\Sigma(\omega=\epsilon_n^{QP}) - v_{xc}
\right|
\bar{\psi}_n
\right\rangle ,
\label{eq:GW_QP_EQ}
\end{equation}
where the OPAW-DFT orbitals $|\bar{\psi}_n\rangle$ are used in place of the true Dyson orbitals.

The sGW approach employs the Hedin space–time formulation of the self-energy  \cite{Hedin1965},
\begin{equation}
\Sigma(r,r',t) = i\, G(r,r',t)\, W(r,r',t^+),
\label{eq:self_ene}
\end{equation}
where $G$ is the single-particle Green's function and $W$ is the screened Coulomb interaction. The first step is to stochastically sample the Green's function $G(t)$. 
We begin by defining the projector onto the occupied KS manifold,
\begin{equation}
\hat{P}
=
\sum_{n \leq N_{\mathrm{occ}}}
|\bar{\psi}_n\rangle \langle \bar{\psi}_n| .
\label{eq:projector}
\end{equation}
The zero-order Green's function $G_0(t)$ is
\begin{equation}
iG_0(t)
=
e^{-i \bar{H}_0 t}
\left[
(1-\hat{P})\,\theta(t)
-
\hat{P}\,\theta(-t)
\right],
\label{eq:zero_G}
\end{equation}
and one approximates $G \approx G_0$. Here $\bar{H}_0$ is the OPAW KS-DFT Hamiltonian and $\theta(t)$ is the Heaviside step function. 

To evaluate $G(t)$, we introduce a set of random orbitals $\bar{\zeta}(r)$ whose components are chosen as $\pm 1/\sqrt{dV}$, where $dV$ is the volume element of the real-space grid. It can then be shown that the Green's function can be written as:
\begin{equation}
iG_0(r,r',t)
\approx
\frac{1}{N_{\bar{\zeta}}}
\sum_{\bar{\zeta}}
\zeta(r,t)\,\bar{\zeta}^*(r'),
\label{eq:zero_G_stochastic}
\end{equation}
where $N_{\bar{\zeta}}$ is the number of stochastic samples (Monte Carlo realizations) while the time-dependent stochastic orbitals $\zeta(r,t)$ are separated into negative-time (hole) and positive-time (electron) components. For negative times, the stochastic orbital is
\begin{equation}
\zeta(r,t<0)
=
iG_0(t<0)\bar{\zeta}(r)
=
-e^{-i\bar{H}_0t}\hat{P}\bar{\zeta}(r),
\label{eq:zero_G_occ}
\end{equation}
while for positive times it is
\begin{equation}
\zeta(r,t>0)
=
iG_0(t>0)\bar{\zeta}(r)
=
e^{-i \bar{H}_0 t}
(1-\hat{P})\bar{\zeta}(r).
\label{eq:zero_G_val}
\end{equation}

In practice, the statistical error of the self-energy and GW corrections is determined by $N_{\bar{\zeta}}$. Achieving an accuracy of roughly 1–2 kcal/mol requires a large number of samples for small systems (on the order of thousands), making the stochastic approach inefficient in this regime. For larger systems, however, fewer samples are required; for systems with tens or more electrons, typically $N_{\bar{\zeta}}\approx500$ is sufficient due to self-averaging.

Given the stochastic expansion $G(t)\approx G_0(t)$, the self-energy matrix element becomes
\begin{equation}
\langle \bar{\psi}_n | \Sigma(t) | \bar{\psi}_n \rangle
=
\frac{1}{N_{\bar{\zeta}}}
\sum_{\bar{\zeta}}
\langle \zeta^*(t)\bar{\psi}_n | W(t) |\bar{\zeta}\bar{\psi}_n \rangle,
\label{eq:self_exp}
\end{equation}
which, using the definition
\begin{equation}
u(r,t)
=
\int
W(r,r',t)
\bar{\zeta}(r')\bar{\psi}_n(r')
dr',
\label{eq:u_TO_pot}
\end{equation}
and the retarded (causal) quantity
\begin{equation}
u^R(r,t)
=
\int
W^R(r,r',t)
\bar{\zeta}(r')\bar{\psi}_n(r')
dr',
\label{eq:retard_pot}
\end{equation}
gives
\begin{equation}
\langle \bar{\psi}_n | \Sigma(t) | \bar{\psi}_n \rangle
=
\frac{1}{N_{\bar{\zeta}}}
\sum_{\bar{\zeta}}
\langle \zeta(t)\bar{\psi}_n | u(t) \rangle .
\label{eq:Wmat_elem_u}
\end{equation}
This converts the difficult direct-product self-energy matrix element into an expectation value of the time-ordered polarization operator $W(t)$ evaluated over $N_{\bar{\zeta}}$ stochastic vectors, $\bar{\zeta}$.

The matrix elements of the retarded (causal) effective interaction, $W^R$, can be obtained through time-dependent Hartree (TDH) propagation. Given the source term, which is the “ket” in Eq.(\ref{eq:self_exp}), $\bar{\zeta}(r)\bar{\psi}_n(r)$, one “disturbs” the sea of electrons with the electrostatic potential produced from this source. The subsequent evolution of the induced Coulomb potential yields the action of the retarded interaction $W^R(t)$ on the source term Eq. (\ref{eq:retard_pot}).

The complication is that to follow the dynamics of the electron sea one would need  to represent and propagate all occupied molecular orbitals. However, here again a stochastic method, i.e, stochastic TDH (sTDH), is applicable and reduces the scaling, as we detail below \cite{Neuhauser2014}. One begins first by defining a set of stochastic occupied orbitals, defined as:
\begin{equation}
\bar{\eta}_l(r)=\sum_{i=1}^{N_{\rm{occ}}} \pm \bar{\psi}_i(r), 
    \quad l = 1, \ldots, N_{\bar{\eta}}
    \label{eq:occ_orb}
\end{equation}
where each $\bar{\eta}_l$ is a random linear-combination of the $N_{\rm{occ}}$ occupied orbitals. Only a few stochastic occupied orbitals are needed, with $N_{\bar{\eta}}\approx 8-20$ used in this work, with fewer terms  needed for bigger systems, due to self-averaging. Note that a new set of $\bar{\eta}_l$ is chosen for each of the $N_{\bar{\zeta}}$ stochastic realizations of the Green’s function in Eq.(\ref{eq:zero_G_stochastic}). The number $N_{\bar{\eta}}$ is typically much smaller than $N_{\bar{\zeta}}$, because it describes the plasmon-like response of the electron sea, which is largely classical, unlike the interference-dominated Green’s function in Eq.(\ref{eq:zero_G_stochastic}).
The stochastic occupied orbitals are then perturbed:
\begin{equation}
\bar{\eta}_l^{\lambda}(r,t=0^+)
=
e^{-i \lambda v_{\mathrm{pertb}}}
\bar{\eta}_l(r),
\label{eq:occ_orb_pertub}
\end{equation}
where the perturbing potential is 
\begin{equation}
v_{\mathrm{pertb}}(r)
=
\int
|r-r'|^{-1}
\bar{\zeta}(r')\,
\bar{\psi}_n(r')\,
dr',
\label{eq:pertb_vh}
\end{equation}
with a weak perturbation strength $\lambda$; we typically use $\lambda = 10^{-4}$ $\mathrm{Hartree}^{-1}$. Both perturbed and unperturbed orbitals are then propagated under the TDH Hamiltonian:
\begin{equation}
\bar{H}^{\lambda}(t)
=
\bar{H}_0
+
\bar{v}_H^{\lambda}(r,t)
-
\bar{v}_H^{\lambda=0}(r,t)
+
\bar{D}^{\lambda}(r,t)
-
\bar{D}^{\lambda=0}(r,t),
\label{eq:opaw_ham_t}
\end{equation}
with
\begin{equation}
\bar{v}_H^{\lambda} = \hat{S}^{-1/2} v_H^{\lambda} \hat{S}^{-1/2},
\qquad
\bar{D}^{\lambda} = \hat{S}^{-1/2} \hat{D}^{\lambda}\hat{S}^{-1/2}.
\label{eq:opaw_ham_time_parts}
\end{equation}
The time evolution of the stochastic orbitals is obtained by solving the TDH Schrödinger equation within the OPAW framework, written as
\begin{equation}
i \frac{d\bar{\eta}^{\lambda}_l(r,t)}{dt} = \bar{H}(t) {\eta}_l ^{\lambda}(r,t).
\label{eq:TDSE}
\end{equation}
The stochastic density is constructed as
\begin{equation}
n_{\bar{\eta}}^{\lambda}(r,t)
=
\frac{2 C_{\mathrm{norm}}}{N_{\bar{\eta}}}
\left[
\tilde{n}_{\bar{\eta}}^{\lambda}(r,t)
+
\hat{n}_{\bar{\eta}}(r,t)
\right],
\label{eq:dens}
\end{equation}
\begin{equation}
\tilde{n}_{\bar{\eta}}^{\lambda}(r,t)
=
\sum_{l=1}^{N_{\bar{\eta}}}
\left|
\tilde{\eta}_l^{\lambda}(r,t)
\right|^2,
\qquad
\tilde{\eta}_l^{\lambda}(r,t)
=
\hat{S}^{-1/2}
\bar{\eta}_l^{\lambda}(r,t).
\label{eq:dens_paw}
\end{equation}
Here, $\tilde{n}_{\bar{\eta}}^{\lambda}(r,t)$ denotes the stochastic PAW density, $\hat{n}_{\bar{\eta}}(r,t)$ is the corresponding all-electron correction to the density, and $C_{\mathrm{norm}}$ is a normalization factor that ensures the correct total number of electrons. 

The retarded potential is then obtained from the difference in the perturbed and unperturbed Hartree potentials:
\begin{equation}
u^{R}(r,t) =
\frac{
v_H\!\left[n_{\bar{\eta}}^{\lambda}(r,t)\right]
-
v_H\!\left[n_{\bar{\eta}}^{0}(r,t)\right]
}{\lambda}.
\end{equation}
The time-ordered  $u(r,t)$ should then be constructed by applying the time-ordering operator $\mathcal{T}$ to the retarded potential $u^{R}(r,t)$.

A straightforward time-ordering of the retarded potential is memory intensive. To overcome this, we apply sparse-stochastic compression \cite{Vlek2018}. This general data compression method utilizes sparse, ``fragmented'', vectors, $\{\xi(r)\}$, that each randomly sample a fraction of the full real-space grid. To complete time-ordering of the effective potential, $u(r,t)$, we first project the retarded potential onto the sparse-stochastic basis:  
\begin{equation}
u^R_{\xi}(t)= 
\frac{v_{\xi}^{\lambda}-v_{\xi}^{\lambda=0}}{\lambda},
\quad
v_{\xi}^{\lambda}= \langle \xi| v^{\lambda}_H(t) \rangle
\label{eq:retard_stoc}
\end{equation}

Here, we use an auxiliary basis of $N_\xi=10,000$ sparse vectors to complete the time-ordering, which translates to sparse-vectors of length $L \sim \mathcal{O}(N/N_\xi)$ that map $\mathcal{T}u_{\xi}^R(t)\to u_\xi(t)$, which provides 
\begin{equation}
    u(r,t) \approx \frac{1}{N_{\xi}} \sum_{\xi}\xi(r) u_{\xi}(t).
    \label{eq:nonret_pot}
\end{equation}
For further details on this procedure, see Ref.  \cite{Vlek2018}. 
Finally, gathering Eqs. (\ref{eq:self_ene}), (\ref{eq:projector})-(\ref{eq:retard_pot}), and (\ref{eq:nonret_pot}), yields the working  equation for sGW,

\begin{equation}
\langle \bar{\psi}_n \left| \Sigma(t) \right| \bar{\psi}_n \rangle
\approx
\frac{1}{N_{\bar{\zeta}}}
\sum_{\bar{\zeta}}
\int
\bar{\psi}_n(r)\,
\zeta(r,t)\,
u(r,t)\,
dr.
\label{eq:stoc_self_ene}
\end{equation}
Note that for simplicity the deterministic orbitals are all assumed here to be real, and the same for the stochastic coefficients, $\bar{\eta}(r)$ and $\bar{\zeta}(r)$, at $t=0$.  The formalism easily carries over to the general case.
\section{Results}
We demonstrate that OPAW-sGW achieves an accuracy comparable to that of the previously developed NCPP-sGW implementation. This conclusion is supported by direct benchmarking of OPAW-sGW results against our earlier NCPP-sGW calculations. We first perform convergence studies with respect to grid spacing on a small system, naphthalene. We then apply the OPAW-sGW method to large molecules and compare energies and computational costs relative to NCPP-sGW.

\subsection{Accuracy of OPAW-sGW compared to NCPP-sGW}
\begin{figure}[h]
  \centering
  \includegraphics[width=0.9\linewidth]{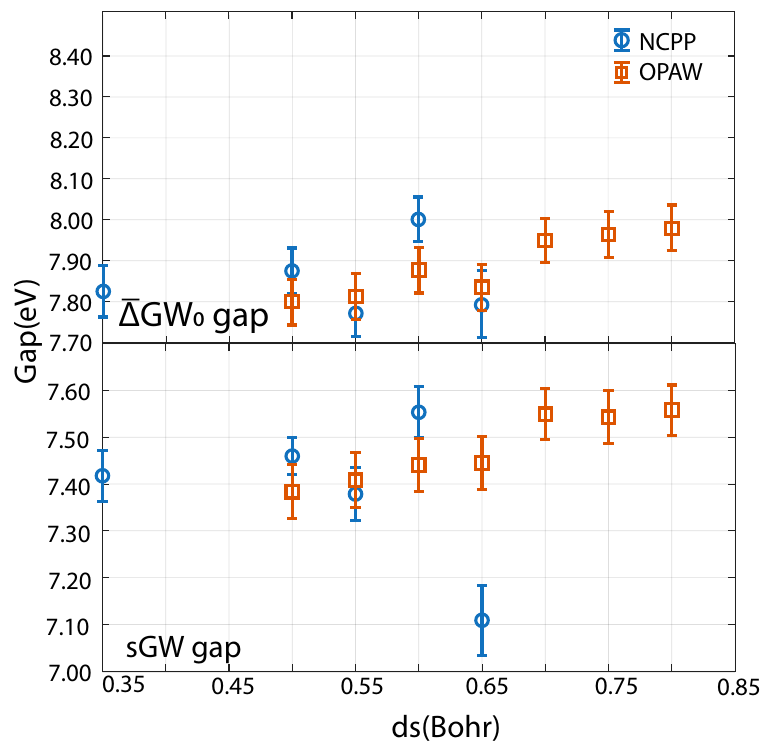}
  \caption{sGW and $\bar{\Delta}GW_0$ QP band gap comparison for naphthalene at varied grid spacings between NCPP and OPAW.}
  \label{fig:naph_gap_comparison}
\end{figure}

\FloatBarrier
\begin{figure}[h]
    \centering
    \includegraphics[width=1\linewidth]{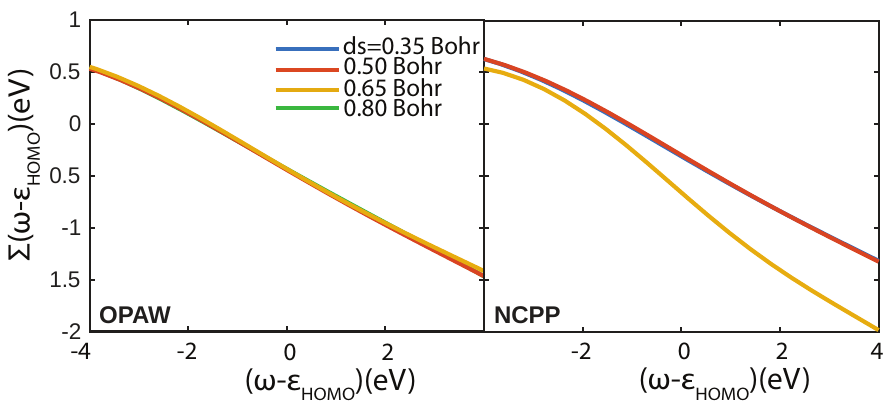}
    \caption{NCPP and OPAW HOMO self-energy plots at varied grid spacings for naphthalene. 
The self-energy is evaluated at the difference between a specific energy and the quasiparticle (QP) energy of the HOMO ($\varepsilon_{\mathrm{HOMO}}$) at the corresponding level of theory and grid spacing.}
  \label{sigma_naph}
\end{figure}

To characterize the effective grid resolution, we define the average grid-spacing quantity $ds = (dV)^{1/3}$. For both approaches, the initial KS wavefunctions were generated using the Chebyshev-filtered subspace method \cite{Zhou2006}, and the local density approximation (LDA) was used for the exchange–correlation potential. Long-range Coulomb interactions were treated with the Martyna–Tuckerman scheme \cite{Zhou2014,Martyna1999}. In the OPAW calculations, LDA atomic datasets from the ABINIT database were used \cite{Jollet2014}, whereas the NCPP calculations employed norm-conserving pseudopotentials in the Hamann form \cite{Hamann1979}. For all simulations, the computational domain was chosen such that the simulation box boundaries were at least $6$ Bohr from the molecule in every spatial direction. Within the OPAW-sGW framework, Eqs.(\ref{eq:zero_G_occ})–(\ref{eq:zero_G_val}) and (\ref{eq:TDSE}) were performed using a fourth-order Runge–Kutta integrator (RK4) with a time step of $0.05$ atomic units. In contrast, the NCPP-sGW calculations used a split-operator propagation scheme with the same time step. The split-operator method applies a Trotter decomposition of the Hamiltonian into kinetic and potential operators, yielding unitary and typically stable time evolution. This approach is well suited for NCPP, where the Hamiltonian separates naturally into these terms. 

In OPAW, however, the Hamiltonian in Eq.(\ref{eq:opaw_ham}) does not admit a straightforward kinetic–potential decomposition. We therefore employ RK4, which is suitable for general non-separable Hamiltonians. At the same time step, however, RK4 requires approximately four times as many Hamiltonian applications as the split-operator method.

To assess convergence with respect to grid spacing, we follow the procedure used in our previous OPAW-TDDFT study \cite{Nguyen2024}. We compute the quasiparticle (QP) band gap of naphthalene over a range of grid spacings for both the NCPP and OPAW approaches. Fig.\ref{fig:naph_gap_comparison} shows the resulting QP band gaps as a function of grid spacing for both methods, with and without the $\bar{\Delta}GW_0$ correction.

For NCPP, the band gap begins to lose accuracy at $ds \approx 0.55$ Bohr and becomes nonphysical beyond $ds \approx 0.6$ Bohr. In contrast, the OPAW band gaps remain stable and quantitatively comparable to the NCPP results at substantially larger grid spacings, up to $ds = 0.8$ Bohr. In particular, the accuracy achieved by NCPP in the range $ds = 0.35$–$0.5$ Bohr is reproduced by OPAW at grid spacings as large as $ds = 0.8$ Bohr.
\begin{figure}[h]
    \centering
    \includegraphics[width=0.9\linewidth]{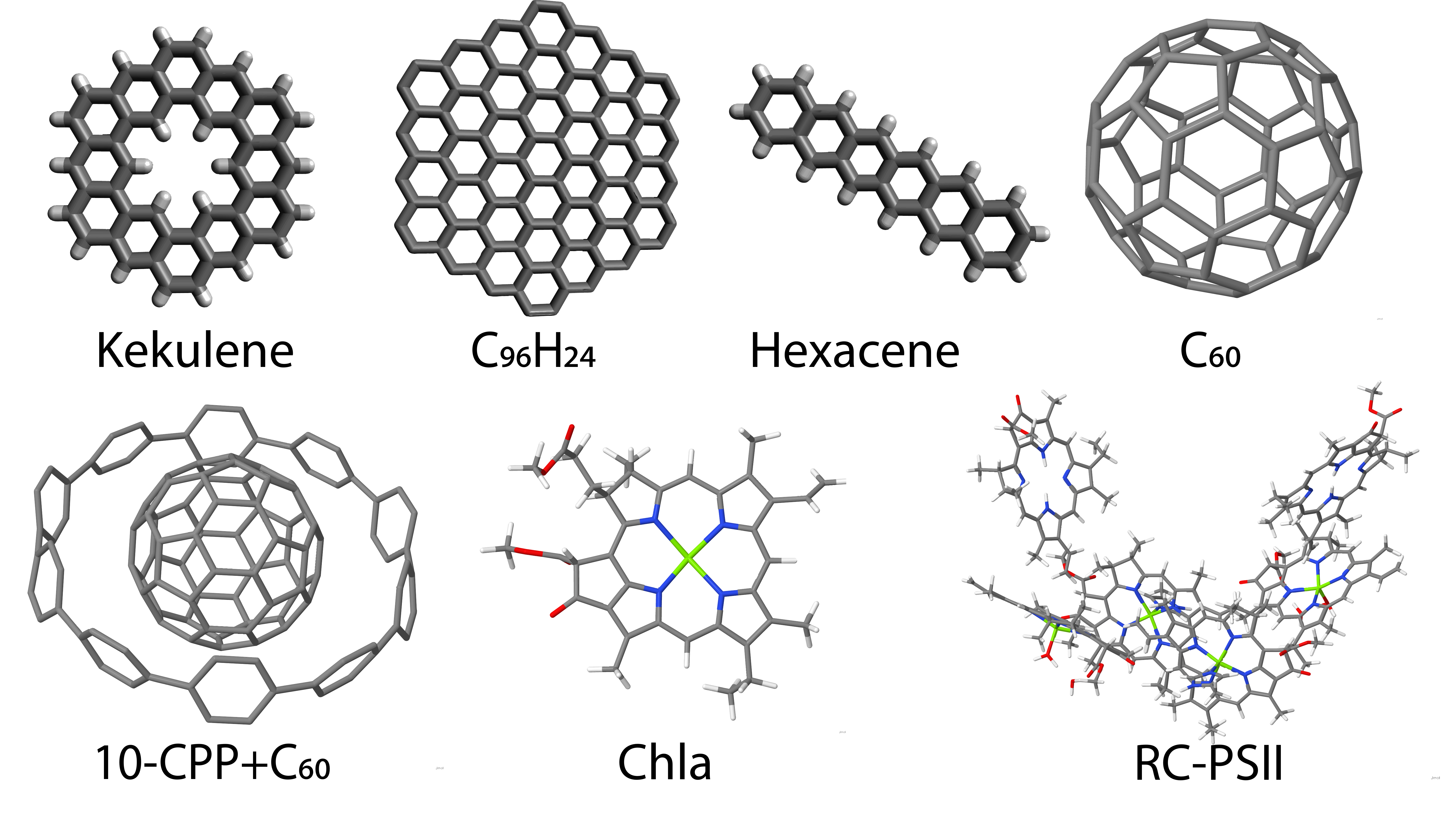}
    \caption{Molecular systems studied in this work.}
  \label{molecules}
\end{figure} 
  
\FloatBarrier

\begin{table}[h]
    \centering
    \begin{tabular}{lcccc}
        \hline\hline
        Molecule
        & $ds$
        & sGW
        & $\bar{\Delta} GW_0$
        & Std. Dev. \\[-2pt]
        & (Bohr)
        & (eV)
        & (eV)
        & ($\pm$ eV) \\
        \hline
        Hexacene                     & 0.77 & 3.63 & 3.84 & 0.06 \\
        Kekulene                     & 0.65 & 4.92 & 5.16 & 0.05 \\
        $\mathrm{C}_{60}$            & 0.79 & 4.33 & 4.58 & 0.05 \\
        Chla                         & 0.70 & 3.71 & 3.86 & 0.06 \\
         $10\text{-CPP}+\mathrm{C}_{60}$& 0.90 & 3.62 & 3.84 & 0.04 \\
        $\mathrm{C}_{96}\mathrm{H}_{24}$ & 0.80 & 3.19 & 3.31 & 0.04 \\
        RC-PSII                      & 0.75 & 3.54 & 3.73 & 0.03 \\
        \hline\hline
    \end{tabular}
    \caption{OPAW QP band gaps calculated using sGW and eigenvalue self-consistent GW
    ($\bar{\Delta} GW_0$) for various molecular systems at the indicated grid
    spacing $ds$. The standard deviation in the QP bandgaps computed using sGW and $\bar{\Delta}GW_0$.}
    \label{tab:opaw_gaps}
\end{table}

\begin{table}[h]
    \centering
    \begin{tabular}{lcccc}
        \hline\hline
        Molecule
        & $ds$
        & sGW
        & $\bar{\Delta} GW_0$
        & Std. Dev. \\[-2pt]
        & (Bohr)
        & (eV)
        & (eV)
        & ($\pm$ eV) \\
        \hline
        Hexacene                     & 0.50 & 3.68 & 3.90 & 0.05 \\
        Kekulene                     & 0.50 & 4.84 & 5.11 & 0.05 \\
        $\mathrm{C}_{60}$            & 0.50 & 4.37 & 4.60 & 0.04 \\
        Chla                         & 0.50 & 3.81 & 3.96 & 0.06 \\
         $10\text{-CPP}+\mathrm{C}_{60}$& 0.50 & 3.63 & 3.85 & 0.04 \\
        $\mathrm{C}_{96}\mathrm{H}_{24}$ & 0.50 & 3.04 & 3.16 & 0.04 \\
        \hline\hline
    \end{tabular}
    \caption{Analogous to Table \ref{tab:opaw_gaps}, but with NCPP and excluding the largest system, RC-PSII.}
    \label{tab:ncpp_gaps}
\end{table}
\FloatBarrier
Our results show that naphthalene converges at a coarser grid spacing with OPAW than with NCPP. Based on this observation, we extend our study to larger systems: hexacene, kekulene, fullerene ($\mathrm{C}_{60}$), chlorophyll \textit{a} (Chla), $10$-CPP$+\mathrm{C}_{60}$, $\mathrm{C}_{96}\mathrm{H}_{24}$, and the Photosystem~II reaction center (RC-PSII). RC-PSII is a dye complex consisting of six chromophores. The atomic coordinates for the RC-PSII system were taken from Ref.~\cite{Frster2022}, where they were optimized at the PBE/def2-TZVP-MM level. The molecular structures of all systems considered are shown in Fig.\ref{molecules}. For all systems except RC-PSII, quasiparticle (QP) band gaps were computed using both NCPP and OPAW. For RC-PSII, NCPP calculations were computationally prohibitive, so the simulations were restricted to OPAW. 

The OPAW and NCPP QP band gaps are summarized in Tables \ref{tab:opaw_gaps} and \ref{tab:ncpp_gaps}. GW band gaps for hexacene, kekulene, $\mathrm{C}_{60}$, Chla, $10$-CPP$+\mathrm{C}_{60}$, and $\mathrm{C}_{96}\mathrm{H}_{24}$ were computed using both NCPP-sGW and OPAW-sGW. 

For all systems except hexacene and RC-PSII, we used $N_{\bar{\zeta}} = 2000$ and $N_{\bar{\eta}} = 8$. For hexacene, $N_{\bar{\zeta}} = 3000$ and $N_{\bar{\eta}} = 16$ were used, while RC-PSII employed $N_{\bar{\zeta}} = 900$ and $N_{\bar{\eta}} = 8$. For the OPAW calculations, grid spacings in the range $ds = 0.65$–$0.9$ were used, whereas the NCPP calculations employed a fixed spacing of $ds = 0.5$. For each molecule, the OPAW value of $ds$ was chosen as the largest grid spacing for which the OPAW DFT band gap differed by only a few meV from the value obtained at the finer grid spacing $ds = 0.5$. At these respective grid spacings, OPAW reproduces the NCPP results obtained with finer grids, with differences ranging from $0.01\,\mathrm{eV}$ to $0.15\,\mathrm{eV}$.

\begin{figure}[h]
  \centering
\includegraphics[width=1.0\linewidth]{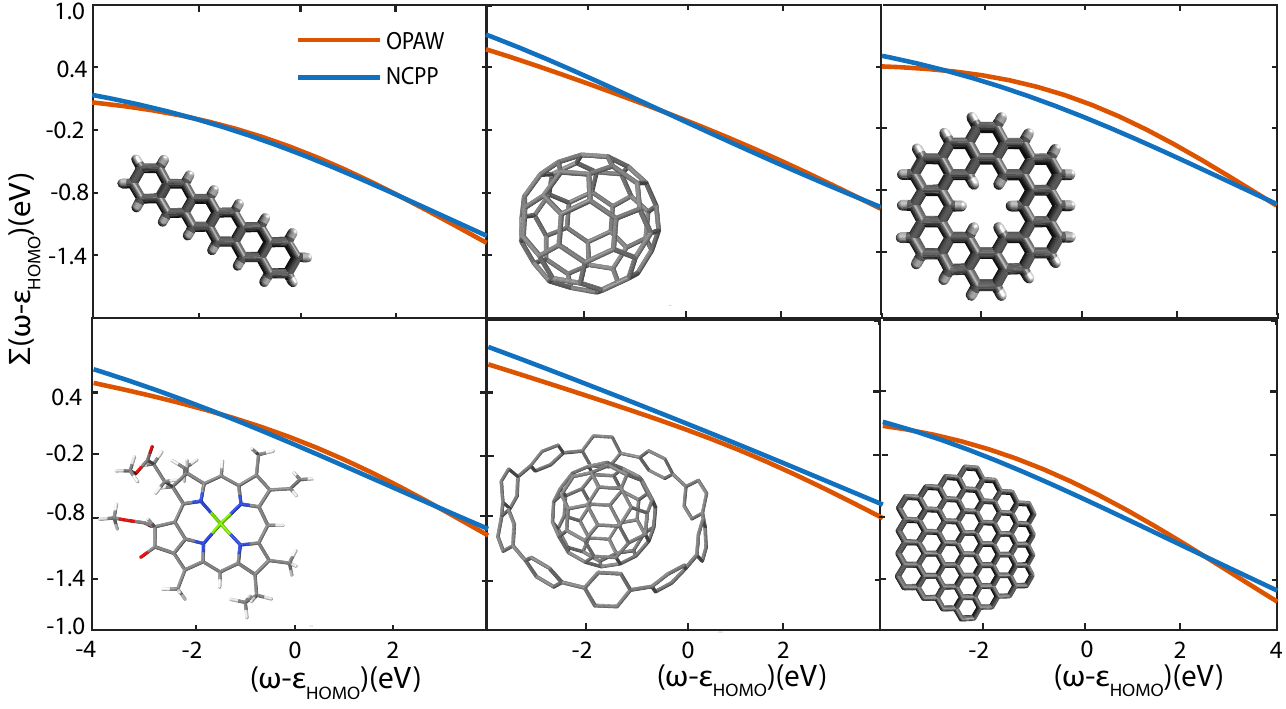}
\caption{HOMO self-energy $\Sigma(\omega)$ calculated for Hexacene, $\mathrm{C}_{60}$, Kekulene, Chla, $10\text{-CPP}+\mathrm{C}_{60}$, and $\mathrm{C}_{96}\mathrm{H}_{24}$. The grid spacings used for the NCPP and OPAW calculations are listed in Tables I and II.The self-energy is evaluated at the difference between a specific energy and the quasiparticle (QP) energy of the HOMO ($\epsilon_{HOMO}$) at the corresponding level of theory, grid spacing, and molecular system.}
\label{fig:molecules_sigma}
\end{figure}

To further illustrate this agreement, we compare the HOMO self-energy curves for representative molecules listed in Tables \ref{tab:opaw_gaps} and \ref{tab:ncpp_gaps}. Fig. \ref{fig:molecules_sigma} shows the self-energies for systems exhibiting the closest agreement in band gaps, including $\mathrm{C}_{60}$ and $10$-CPP$+\mathrm{C}_{60}$, as well as systems with the largest deviations, namely Chla and $\mathrm{C}_{96}\mathrm{H}_{24}$. The trends observed in these self-energy plots are consistent with the band-gap differences reported in Tables~\ref{tab:opaw_gaps} and \ref{tab:ncpp_gaps}.

\FloatBarrier
\subsection{Computational Speed}
Our results show that OPAW-sGW attains an accuracy comparable to NCPP-sGW while operating at substantially larger grid spacings. However, because OPAW-sGW employs an RK4 time-propagation scheme rather than the split-operator approach used in NCPP-sGW, the overall computational speed of the two methods remains similar. As an illustrative example, we compare the wall times for the $10$-CPP$+\mathrm{C}_{60}$ system using $2000$ Monte Carlo samples parallelized over $400$ CPU cores. For OPAW-sGW, using $ds = 0.9~\mathrm{Bohr}$, the total wall time is approximately $3.7$ hours, whereas NCPP-sGW with $ds = 0.5~\mathrm{Bohr}$ requires about $2.6$ hours.

This difference in wall time arises from the time-propagation scheme used in each method. In OPAW-sGW, the RK4 integrator requires four Hamiltonian applications per time step. In contrast, the split-operator scheme used in NCPP-sGW requires only a single Hamiltonian application per step. Attempts to increase the time step further led to numerical instabilities during testing. Unlike our previous OPAW-TDDFT implementation, the RK4 scheme cannot be parallelized over stochastic orbitals because the stochastic orbitals must be generated and propagated independently on each processor core. Although the cleaning procedure introduced in Ref.~\cite{Bradbury2022} mitigates time-propagation instabilities, it does not allow a further increase in the time step.

Despite this limitation, the memory savings provided by OPAW-sGW are substantial. For example, in the case of $10$-CPP$+\mathrm{C}_{60}$, the OPAW grid contains roughly one-sixth the number of points required by NCPP. This significant reduction in memory usage enables sGW calculations on much larger molecular systems.

\section{Conclusion} 
Here we integrate the OPAW formalism into the sGW framework and systematically compare its performance with the established NCPP-sGW approach. Across a diverse set of hydrocarbon and biologically relevant systems, OPAW-sGW reproduces QP band gaps and self-energy spectra that are quantitatively comparable to those obtained with NCPP-sGW using fine real-space grids, while operating at substantially larger grid spacings. 

Although the current OPAW-sGW implementation has a higher CPU cost, even at larger grid spacings, than NCPP-sGW, this is offset by substantial memory savings from using fewer grid points. Moreover, because OPAW retains the full accuracy of the PAW method, as demonstrated in previous work \cite{Li2020}, it produces highly accurate electronic states near the atomic cores. This enables more reliable benchmarking relative to NCPP approaches, which rely on smoothed core potentials.

Given these capabilities for electronic structure calculations within the OPAW framework, we outline several future directions:

First, OPAW-DFT will be combined with mixed deterministic/sparse-stochastic compression of exchange integrals \cite{Bradbury2023_2} to enable efficient generalized Kohn–Sham (GKS) calculations, i.e., DFT including a fraction of exact exchange. This approach benefits from a static grid-to-basis transformation that takes advantage of the higher grid resolution available in OPAW.

Second, the efficient evaluation of matrix elements of the screened Coulomb interaction $W$ within OPAW-sGW enables the construction of statically screened Coulomb matrices required for BSE spectra using the same stochastic TDH approach \cite{Bradbury2022,Bradbury2023,Li2025,Bradbury2024}. The accurate treatment of core states in OPAW further enables the calculation of core-level excitation spectra within both TDDFT and BSE using stochastic methods, extending previous deterministic studies \cite{Vinson2010,Ljungberg2011}.

Finally, OPAW-sGW can be extended to incorporate vertex corrections to the self-energy, as previously applied to one-shot sGW in Ref.~\cite{Voj}. In the sGW framework, low-order vertex corrections correspond to a modified time-dependent Hartree–Fock (TDHF) propagation rather than TDH propagation to obtain the polarization potential \cite{Romaniello2009,Schindlmayr1998}. Because OPAW permits coarser real-space grids, it improves the computational efficiency of calculations involving vertex-corrected GW schemes.
\section{Acknowledgments}
This work is supported by the National Science Foundation (NSF) under Grant No. CHE-2245253. Computational resources were provided by the Expanse Cluster at the San Diego Supercomputer Center through allocation PHY240131 under the Advanced Cyberinfrastructure Coordination Ecosystem: Services \& Support (ACCESS) program.

\section*{Declaration of Competing Interest}
The authors declare that they have no known competing financial interests or personal relationships financial interests or personal relationships that could have appeared to influence the work reported in this paper.

\section*{Data Availability Statement}
The data that support the findings of this study are available from the corresponding author upon reasonable request.
\bibliographystyle{achemso}
\bibliography{ref}

\end{document}